# Semi-automatic Assessment Model of Student Texts - Pedagogical Foundations

*Tuomo Kakkonen, Erkki Sutinen*

University of Joensuu, Finland

**Abstract:**

> *This paper introduces the concept of the semi-automatic assessment of student texts that aims at offering the twin benefits of fully automatic grading and feedback together with the advantages that can be provided by human assessors. This paper concentrates on the pedagogical foundations of the model by demonstrating how the relevant findings in research into written composition and writing education have been taken into account in the model design.*

## 1   Introduction

The authors of the current paper have coined the term *semi-automatic assessment* to describe the features of the next generation assessment model for student texts, which is designed to offer the benefits of both fully automatic grading and feedback as well as the advantages that are gained when human assessors (the teacher, the student him/herself and peers) make assessments on the basis of their personal judgments as well as on the basis of the evidence collected by automatic assessment components. Such an undertaking needs to take into account linguistic, technical and pedagogical research findings. While the technical and linguistic aspects of the semi-automatic assessment model have been already explored in some detail in our earlier publications, the pedagogical foundations have not yet been sufficiently investigated. This paper intends to fill that gap. We review the research findings about written composition and writing education, and incorporate the best recommendations from that research into the semi-automatic assessment model.

The paper is organized as follows. Section 2 introduces the background to this work and outlines the concept of semi-automatic assessment. Section 3 sets out the results of the investigation of the relevant literature on writing and writing assessment. Section 4 explains how we combined the best practices in order to produce the semi-automatic assessment model. Section 5 concludes our findings.

## 2   Background

### 2.1   Automatic grading and feedback

A large number of studies (see, for example, [1]) have reported on automatic essay grading experiments in which the system has been able to equal and even surpass the levels of agreement between the grades assigned by human graders. The most obvious motivation for developing fully automatic grading systems is the enormous savings that such systems could effect in assessment costs. Automatic grading also contributes to the objectivity and reliability of grading because a grading system is not affected by purely human factors such as boredom and fatigue.

While the automatic grading of essays is not a new idea, and despite the impressive results achieved by current state-of-the-art systems, automatic graders are not yet being widely utilized in real-life teaching situations. Article [1] offers a good summary of the various criticisms that have been leveled against the automatic assessment of student writing. Huot and Neal [2], for example, point out that the purely *technocentric* understanding of writing assessment has resulted to identifying the main problem associated with the assessment of student texts as "the inability of readers to give the same scores to the same papers".

Various systems that are currently being developed are capable of providing assessments that go beyond mere holistic grading [3]. The best of these systems allocate separate grades for other aspects of essay writing such as spelling and grammar. Some systems have also been designed to provide learners with instructions about how to improve their text. Such feedback might, for example, include comments on how to improve the grammatical correctness of the text. In spite of the large number of successes that researchers have achieved in improving the performance of automatic assessment systems, this field of research is still beset by formidable challenges. While it is indeed now possible to provide a substantial amount of automatic feedback, the pedagogical utility of such feedback is often limited in its effectiveness because it concerns itself almost exclusively with fundamental writing skills such as spelling mistakes and grammatical errors.

### *2.2  Semi-automatic assessment*

While automated feedback is certainly a step in the right direction, our argument in this paper is that the design of current assessment systems places far too great an emphasis on technocentric considerations rather than on the achievement of sound pedagogical principles. The reasons for this are obscure, but one may speculate that they arise because of a lack of collaboration between those who work on developing the technology itself and those who are concerned with implementing the concomitant pedagogical aims of educational research. We share the views of, for example, [2, 5], that while technology can indeed make a vital contribution to the assessment of writing and the provision of feedback, the pedagogical aims and rationale, and the ways in which technologies can be applied to achieve these aims, needs to be much more thoroughly investigated than has been done in the past.

Our response to this need has been to design and develop the concept of the semi-automatic assessment of student texts, which radically departs from the usual assumptions about the principles that govern the practice of both the manual (human) and automatic assessment of student writing. The most obvious advantage of manual assessment over automatic assessment is that it can produce a high level of personalization. A teacher who is familiar with the strengths and weaknesses of a particular student will be able to tailor his or her feedback to meet the needs of the student by supplying individualized comments and advice. Fully automatic grading cannot take the personal needs and expectations of a student into account in the way that another human being can do. But assessment and the compilation of precise and comprehensive feedback is not an easy task, and it is in this respect that computers can support teachers by offering automatic support and "evidence" on which to base assessment decisions and feedback.

Semi-automatic assessment permits both the students and the teacher to take part in the assessment processes. Instead of assessing only the *product* (i.e. the final text), a semi-automatic system creates an environment that supports the actual *processes* of writing (composition) and assessment. This kind of assessment incorporates both quantitative or numerical assessments (numerical grades), in addition to qualitative feedback (written feedback, evaluations and instructions). What is achieved by the semi-automatic assessment model is that it allows students, their peers and their teachers to process and analyze a text by identifying its strong points and shortcomings, to check for possible instances of plagiarism, and to provide comments and instructions in a variety of formats. The EssayAid system that we have been developing over the past seven years represents a valuable contribution towards the ultimate realization of a functional semi-automatic assessment system.

# 3   Review of relevant pedagogical research

## 3.1   The role of technology in the assessment of student writing

Many researchers of writing instruction are of the opinion that technology can and will in the future play an important role in the assessment of written compositions. While Troia [4], for example, recognizes the need for providing "immediate, instructionally relevant multi-vector data to teachers so that they are better equipped for pinpointing writing problems and responding accordingly", he notes that none of the existing models is adequate for the task. Smith's [6] observations also seem to support models such as semi-automatic assessment. Instead of devoting his research resources to the problem of the reliability of scoring, Smith has been focusing on the *adequacy of the decisions* that are the culmination of the assessment process. Huot and Neal [2] have noted, in reference to this work, that "Smith's research frames the problem for writing assessment as the setting of an appropriate context within which teachers can read students' writing and make informed decision about them".

## 3.2   Product vs Process Models of Writing and Summative vs Formative Assessment

In *product-oriented* models of written composition, an instructor is defined as a spotter of errors and reinforcer of rules [7]. In terms of this model, writing is defined as the process of being able to record preprocessed and fully formed ideas. While this type of assessment is quick to perform, the negative feedback that it emphasizes can be very discouraging and demoralizing for learners. While the focus of research into written language composition shifted several decades ago from product to process, automatic assessment procedures have not followed this trend. Process-oriented writing is based on research into how "real-world" writers compose texts [9]. It is widely accepted that such skills are the very ones that students should learn in schools. While there no single definition of a *process writing* model that is acceptable to all researchers, the definitions that are available in the literature have various features in common [7, 8, 9, 10]:
- The emphasis is more on the writing process than on the final product.
- The importance of the content of the text (i.e. the ideas and meanings that are expressed by the writing) are given precedence over concern about the form in which they are expressed (i.e. the correctness of spelling and grammar).
- The process model strives to be learner-centered rather than teacher-centered. The priority is therefore to help students to identify their problems.

- The process model emphasizes the social implications of writing by according importance to the role of the envisaged audience and the interactions that take place between the reader and the writer.

Models of process writing differ widely on the actual sub-processes that each of them define. Most of the available models at least make a distinction between planning, writing and revising. The sub-processes are also understood to operate recursively [8, 10]. Planning, for example, can take place during revision.

*Summative assessment* is based on evaluating the learning achievements after the learning process itself [6]. Summative assessment is therefore very similar to what happens in the product model of writing. This kind of assessment has been widely used for comparing learning outcomes over a period of time and also for comparing the achievements of educational institutions, the efficacy of particular teaching methods, and so on. This approach to assessment has frequently been criticized for its emphasis on the rating and ranking of students rather than on to maximizing their learning outcomes and providing support for students while they are engaged in the actual learning process. Despite its evident shortcomings, summative assessment is often used to assess student texts. It might, for example, be impractical to use any other form of assessment in the context of examinations because of limitations on resources and time.

*Formative assessment*, by contrast, evolves as a result of a series of interactions between the learner and the educator, and it occurs as a result of (1) the feedback that the learner receives during the learning process and (2) the learner's own engagement in self-reflective contemplation of his or her own efforts to improve. Formative assessment is therefore very similar to what happens during the application of the process model of writing skills acquisition.

### 3.3 Feedback and revision

*Feedback* is designed to help students to improve the quality of their writing [11]. Product-oriented models of composition regard "feedback" as providing the student with a catalogue of errors. The process writing model encouraged to shift the focus of feedback onto higher level of language knowledge, such as content and organization. At its best, formative feedback on writing helps learners to become aware of the gaps in their knowledge. Information of this kind is what a student needs to improve his or her current text and contributes to the eventual realization of ultimate goal of improved writing skills.

Feedback works by initiating student's self-assessment (see Section 3.4) and so to revise the text accordingly. In light of research findings (such as those in [10]), the following conclusions can be drawn. Feedback should…
- focus on content, organization and other higher levels of language skills.
- guide students on how to make improvements, and it should not be dedicated solely to the cataloguing of errors.
- be specific and should explain exactly why a particular aspect of the writing is good or unacceptable.
- stimulate learners to think on their own rather than provide the right answers.

Studies on text *revision* often agree that while less skilled writers concentrate mainly on low-level error corrections and rewording, the revisions of skilled focuses on the revision and recasting of ideas and on the logical coherence [9]. The main pedagogical recommendation that emerges from these observations is that the students should be guided towards making revisions of the latter type.

### 3.4  Self- and peer assessment

The rationale for *self-assessment* is that students should be able to analyze their own texts and present such analyses as part of the assessment process for the consideration of the teacher. This empowers the learning process because it requires students to reflect on their own writing processes. In addition to providing feedback and a sense of audience for the writer, *peer assessment* is useful because it coaches students in critiquing skills and strengthens their ability to read texts analytically and critically [9]. Peer assessment inevitably also helps students to become more aware of the problems that they encounter in their own writing. Bloom [12] insists that critiquing should not be thought of as negative and judgmental but rather as creative and supportive. He also points out that it stimulates the ability of a student to use his or her highest order thinking skills. Peer assessment also supports the process model of writing because it emphasizes the social dimensions of writing as an activity [9].

### 3.5  Rubrics

A *rubric* is a grading tool that simplifies the grading process by dividing it into smaller subtasks. Rubrics were originally developed for expediting manual assessment and as a means for promoting high levels of consistency between graders. A rubric consists of a list of the assessment criteria for a specific assignment and descriptions of how to identify different levels of quality [13].

While rubrics were originally designed as grading aids, *instructional rubrics* are designed to offer feedback, and they have the following properties: [13]: 1) They are written in language that is readily comprehensible to the average student. 2) They define and describe high-quality work. 3) They focus on the common errors and weaknesses that appear in students' work, and indicate how they can be avoided. 4) They can be used also for self-assessment.

## 4  Fitting Our Findings into the Semi-automatic Assessment Model

A semi-automatic assessment system for student texts is designed to create an environment that makes it much easier for a teacher to make useful and accurate assessment decisions because it is supported by computer-generated grading and feedback. This approach supports Smith's notion of the adequacy of assessment decisions. We refer to this aspect of the semi-automatic assessment model as *computer-assisted assessment and feedback generation* (CAFG).

The distinction between product and process models of writing, and formative and summative assessment, provide the theoretical basis for the semi-automatic assessment model. While process writing may function as an invaluable tool for young learners who are practicing the skills of composing and for second language learners, it needs to be accepted that the process writing model is not applicable to all situations. It would not be applicable, for example, to the assessment of examination papers and student admission tests because what is needed in those contexts is merely a reliable ranking of students according to predetermined criteria. While both of these models are pedagogically and psychologically well grounded, they should be regarded as complementary rather than rival models. Whereas the product model provides the basis for making final assessments in the semi-automatic assessment model, feedback mechanisms of the kind described above support the process model (and, in particular, the latter phases of the process, namely revision).

We have therefore divided the semi-automatic model into two sub-models: *the summative feedback cycle* and *formative product assessment*. While feedback and revisions are often repeated several times during the process of creating a text, the final formative assessment is not a recursive process. An instructor could use either or both of these sub-models to meet the needs of a specific assessment task. When assessing an essay examination, for example, a teacher would use only the formative product assessment model to allocate final grades for the students' essays. On the other hand, a language teacher might well decide not to use the product assessment but utilize multiple iterations of the summative feedback cycle. But both of these sub-models could be used, for example, for teaching students how to write project reports and theses.

Both of the sub-models can, moreover, be configured to suit an instructor's preferences. They can, in fact, be configured to combine any combination of fully automatic, semi-automatic and/or peer and self-assessment phases. A typical formative assessment of essay answers from an examination would require the generation of both grading and feedback, and this could be achieved by combining automatic feedback with manual (human) scrutiny (CAFG). It might be ideal to assess a project report document by directing it through two feedback cycles that consist of computer-generated feedback about grammar and vocabulary, peer assessment of the organization and content of the text, and the teacher's comments on the overall organization and coherence of the text. Figure 1 illustrates this idea.

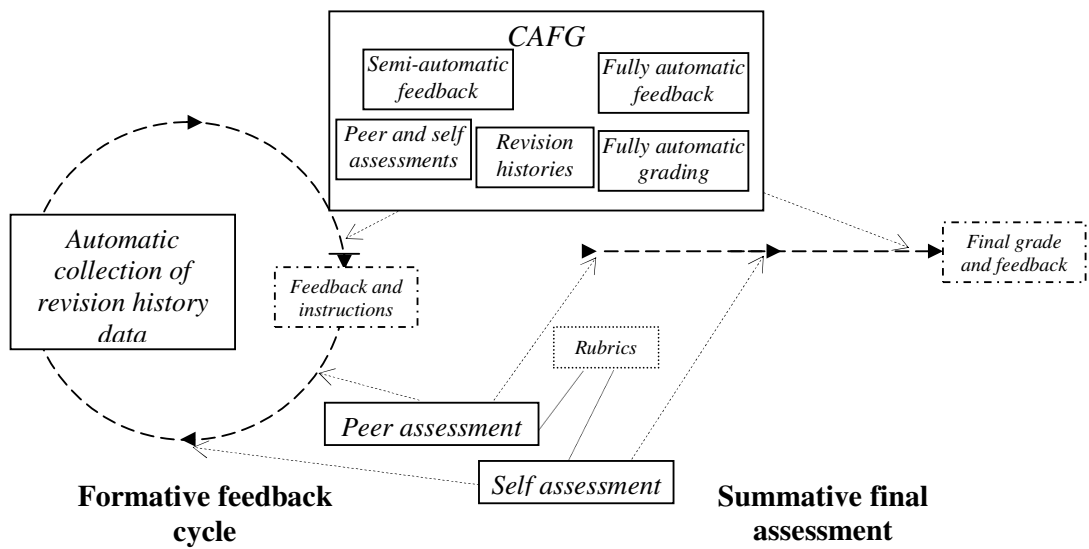

**Figure1**. The two sub-models of semi-automatic assessment.

Figure 1 demonstrates that the CAFG component forms that could be characterized as the "core" activity of the semi-automatic assessment model. It is usually the last phase of an assessment or feedback process. This enables the information collected during the other phases of the assessment to be utilized for helping the teacher to make the final assessment decisions. The revision history can reveal, for example, whether the writer has made only surface-level changes or modified the content and the organization of the text. The ideal scenario is that peer-initiated feedback offers students opportunities to think about new solutions to their problems and to consider new theories. Because the construction of this kind of feedback presupposes a deep understanding of students' texts as well as world knowledge, it is something that is well beyond the capabilities of even the most advanced automatic assessment methods.

While the exact definition of the CAFG component will be left for future work, we present an outline below of those parts of the CAFG that can be implemented on the basis of functionalities that are either already available in EssayAid or that are currently under development:
*Fully automatic grading*. The Automatic Essay Assessor (AEA) is a system for automatically grading essays [14]. The teacher can use the grades assigned by the system as a quality control mechanism by comparing the grades that he or she has assigned to the grades assigned by the system.
*Fully automatic feedback*. The AntiPlag plagiarism detection system enables an instructor to identify those parts of student texts that have been plagiarized from the work of other students works or from the Internet. The RefMan tool that we are currently developing enables instructors automatically to identify (and verify) the citations and references that appear in student texts. This information can be used to automatically identify both extra or missing references and citations. EssayAid also has a component that offers feedback about grammar usage.
*Semi-automatic feedback*. Our TexComp tool is capable of producing comments on the writing style of a text on the basis of its degree of complexity. This tool uses readability and lexical diversity measures to make judgments about the complexity of texts. In its current state, it is not able to provide instructions or suggestions for improvements. But TexComp can be used to alert a teacher to the fact that while one text may exemplify an exceptional degree of complexity, another text may reveal a level of complexity that is far below what is expected from a student who is studying at a particular educational level.

## 5 Conclusion

We have described the semi-automatic assessment model for student texts and reviewed contemporary research on written composition and the assessment of writing. We then combined what we had learned from this analysis in order to create the semi-automatic assessment model which accomplishes the essential pedagogical purposes for which this model was designed. We also defined two sub-models of semi-automatic assessment. By adapting the model to reflect specific combinations of summative feedback cycle and formative final assessment, an instructor may configure the system to offer whatever degree of flexibly an instructor may need for particular assessment tasks. The present model describes the basis on which future versions of the EssayAid assessment system will be built.

In many ways, this work represents only the beginning of ongoing investigations into the pedagogical basis of semi-automatic assessment. In addition to establishing the exact structure of the CAFG component, there are numerous other research questions that need further investigation. Some of the most important research questions that remain to be solved include the following:
*Sub-model configurations*. How can different writing formats (such as theses, project reports, essays, language learning exercises, and so on) be supported by semi-automatic assessment?
*Rubrics*. Instructional rubrics play a key role in peer and self-assessment in the model. Students can be guided to make comments about specific aspects of the evaluated texts by making use of standardized sets of questions. Another way in which rubrics can be used involves completing the final assessment rubrics by making use of both automatic and manual work.

*Revision histories*. How can accumulations of revision histories best be utilized during feedback and in the final assessment? We need to define, for example, how different kinds of revision could be classified and identified.

*Implementation and evaluation.* In order to evaluate the effectiveness of the semi-automatic assessment model, it has to be fully integrated with the EssayAid system. Once this has been done, the effectiveness of the model can be tested in both laboratory and classroom, and the model can then be modified on the basis of the findings.

## Author(s):

Tuomo Kakkonen, PhD
Erkki Sutinen, professor, PhD
University of Joensuu, Department of Computer Science and Statistics
Department of Computer Science and Statistics, University of Joensuu
PO Box 111, 80101 Joensuu, Finland
{tuomo.kakkonen, erkki.sutinen}@cs.joensuu.fi